\documentclass[journal,twocolumn]{IEEEtran}

\usepackage{blindtext, graphicx}
\usepackage{listings}
\usepackage{graphicx}
\usepackage[font=small]{caption}

\usepackage{color}
\usepackage{amsmath,bm}
\usepackage{amsmath}
\usepackage{amssymb}
\usepackage{algorithm}
\usepackage{algpseudocode}
\usepackage{amsthm}
\usepackage{makecell}
\usepackage[utf8]{inputenc}

\makeatletter
\let\NAT@parse\undefined
\makeatother
\usepackage{hyperref}

\title{A Contextual Bandit Approach for Value-oriented Prediction Interval Forecasting }

\author{
\IEEEauthorblockN{Yufan Zhang,  Honglin Wen, \textit{Member, IEEE},  and Qiuwei Wu, \textit{Senior Member, IEEE}}

\vspace{-2em}
\thanks{Yufan Zhang is with the Department of Electrical and Computer Engineering, University of California San Diego, San Diego, California 92161, US. 

Honglin Wen is with Department of Electrical Engineering, Shanghai Jiao Tong University, Shanghai 200240, China.

Qiuwei Wu is with Tsinghua-Berkeley Shenzhen Institute,Tsinghua Shenzhen International Graduate School, Tsinghua University, Shenzhen 518055, China.

Corresponding author: Qiuwei Wu (e-mail: qiuwu@sz.tsinghua.edu.cn).
}
}

\begin{document}

\maketitle
\thispagestyle{empty}
\pagestyle{plain}

\begin{abstract}

Prediction interval (PI) is an effective tool to quantify uncertainty and usually serves as an input to downstream robust optimization. Traditional approaches focus on improving the quality of PI in the view of statistical scores and assume the improvement in quality will lead to a higher value in the power systems operation. However, such an assumption cannot always hold in practice. In this paper, we propose a value-oriented PI forecasting approach, which aims at reducing operational costs in downstream operations. For that, it is required to issue PIs with the guidance of operational costs in robust optimization, which is addressed within the contextual bandit framework here. Concretely, the agent is used to select the optimal quantile proportion, while the environment reveals the costs in operations as rewards to the agent. As such, the agent can learn the policy of quantile proportion selection for minimizing the operational cost. The numerical study regarding a two-timescale operation of a virtual power plant verifies the superiority of the proposed approach in terms of operational value. And it is especially evident in the context of extensive penetration of wind power.


\end{abstract}

\section{Introduction}

The ongoing decarbonization effort in the energy sector places a particular emphasis on renewable energy sources (RESs). Albeit enjoying the merits of clean and non-emission, the stochastic nature of RESs poses a great challenge to power systems operation and electricity markets, as the power generation of RESs cannot be scheduled at will. This drives the need of forecasting RES generation at future times to support power system operation \cite{roald2022power}, such as power dispatch, trading \cite{8352041}, and reserve procurement. 

Forecasts can be communicated in various forms\cite{2020Hong}, including single-valued points\cite{8299478}, densities\cite{wen2022continuous,ZHANG2020105388}, and prediction regions\cite{8253871,5451173}. Among them, prediction regions provide a summary of the probability distribution of random variables. For univariate forecasting, a prediction region is communicated as a prediction interval (PI), which is specified by two bounds and the nominal coverage probability (NCP) $(1-\beta)\times100\%$ that specifies the probability that the realization falls in. PI has a wide range of applications in nowadays power industry. For instance, PI serves as an input to robust optimization for quantifying the wind uncertainty, determining the reserve quantities \cite{zhao2021operating} and wind power offering in the day-ahead market \cite{zhao2021cost}, where the NCP is commonly chosen between 90\% and 95\%. Also, based on the estimated PI, the concept of uncertainty budget is leveraged to reduce the conserveness of robust optimization in storage control \cite{attarha2018adaptive}, unit commitment \cite{bertsimas2012adaptive}, and microgrid dispatch \cite{qiu2018interval}, which is beneficial to reducing the operational cost in the robust optimization.

PI is always desired to have good reliability and sharpness, which means that the interval width needs to be minimized in the constraint of some NCP. In recent decades, non-parametric approaches have been preferred by the forecasting community, which mainly develops quantile regression (QR) models to issue a pair of quantiles as a PI. Machine learning models, such as recurrent neural network \cite{wang2019probabilistic}, ridge regression \cite{zhang2020admm}, and neural basis expansion model \cite{wen2021probabilistic} have been combined with QR, with the loss function of the pinball loss, which shows superiority thanks to the strong learning ability of machine learning models.
Usually, the quantiles in a PI are statistically symmetric with respect to the median, i.e., $q^{\beta/2}$,$q^{1-\beta/2}$, which is therefore referred to as the central PI (CPI) in literature.

However, the probability distribution of the RESs power output is generally skewed \cite{zhao2021cost,4530750}, thereby the width of CPIs is often unnecessarily wide \cite{zhao2020optimal}. For this reason, optimal PI (OPI) forecasting approaches arise, which optimize over the bounds with the objective of improving statistical quality, such as minimizing the Winkler score. A thread of studies select the probability proportion according to the contextual information, instead of setting it as a predetermined constant like in the CPI approaches. Ref. \cite{zhang2022optimal} learned the policy of proportion selection seeking to minimize the Winkler score. In another thread of studies, the forecast model outputs the two bounds directly without specifying a probability proportion to it. As the quality metrics such as Winkler score are generally non-differentiable, the main difficulty lies in how to design a surrogate loss function \cite{pearce2018high} or a proper optimization technique to estimate the forecast model parameters. Ref. \cite{wan2013optimal} formulated a multi-objective problem and optimized the parameters of the extreme learning machine (ELM) by the particle swarm optimization. In \cite{zhao2020optimal}, the parameters estimation for the ELM was formulated as a mixed-integer linear programming (MILP) problem, which was solved by off-the-shelf solvers.

Although the aforementioned PI forecasting approaches have contributed to improving forecasting quality in the view of statistics, they have overlooked the value of forecasts in the downstream power system operation. The idea of using value for evaluating the goodness of forecasting can be dated back to \cite{murphy1993good}, where value is defined as the economic/operational gain from leveraging forecasts at decision-making stages.
Take a robust optimization problem as an example (such as robust power dispatch); the input PI will definitely impact the operational cost. Indeed, it has been shown that the improvement in forecast quality does not necessarily lead to a higher value in operation. For example, the biased prediction of wind power offering quantity is more preferred than the accurate single-point forecasts with small mean squared errors \cite{morales2013integrating,9716858,8706264}, as the operational cost of up-regulation (the case that the wind power offer determined in the day-ahead market is larger than the wind realization) for settling the energy deficit is more expensive than that of down-regulation (the case that the wind power offer determined in the day-ahead market is smaller than the wind realization) for settling the energy excess. Similar results can be found in the context of unit commitment (UC) \cite{chen2022feature}. 

Therefore, value-oriented forecasting has been advocated in recent years \cite{pp,NIPS2017_3fc2c60b}. The key challenge lies in how to link forecasting with decision-making. Attempts have been made by designing decision-aware loss functions. For instance, to fill the gap between the point forecast and decision-making, the loss function called smart "Predict, then Optimize" (SPO) loss was proposed in \cite{elmachtoub2022smart}. Ref. \cite{zhang2022cost} approximated the objective function based on the historical data. However, the approximation may result in errors which may compromise the value of forecasting. In \cite{zhao2021cost}, a cost-oriented machine learning (COML) framework was established, which performed value-oriented PI forecasting by optimizing the probability proportion of QR models under the decision-making objective. However, the COML framework restricts the QR model to be linear so that the estimation of the parameters is allowed to be reduced to a single-level optimization problem through the KKT condition. Here, special focus is placed on multi-timescale decision making in power systems, such as market clearing or centralized operation of virtual power plants (VPPs). As the compensation cost for the decision in each timescale differs, it calls for more strategic forecasting to reduce the cost.

In this paper, without loss of generality, we design a PI forecasting approach for a two-timescale VPP operation task with wind power, where the day-ahead problem is based on robust optimization with recourse, while the real-time problem settles the wind power deviation. Concretely, the proposed value-oriented PI forecasting approach contains a policy learning module, which optimally selects the probability proportion to reduce the operational cost. For that, the training stage of the value-oriented PI forecasting, which involves the estimation of model parameters, is solved by a contextual bandit in a closed-loop manner. Specifically, the policy learning task is modeled by an agent, whereas the optimization over QR models parameters is solved in the environment. The agent and the environment are linked by the reward, which is the negative objective value of the decision-making problem. As such, the agent can learn the selection policy guided by the optimal objective of the decision-making problem. And, the nature of the contextual bandit avoids the tedious work of labelling in supervised learning \cite{bottieau2021automatic}. Compared with the existing studies, the main contributions of the paper are summarized as follows:



1) A new solution strategy for the value-oriented PI forecasting approach, which uses the contextual bandit framework to link the proportion selection with the operational value of the downstream decision-making problem.

2) An integration 
 of the value-oriented PI forecasting approach with the complex decision-making problem, which involves multiple decision variables and constraints.


The remaining parts of this paper are organized as follows. The illustrative examples to show the necessity of value-oriented forecast are presented in Section \uppercase\expandafter{\romannumeral2}. The preliminaries regarding PI and the two-timescale operation are given in Section \uppercase\expandafter{\romannumeral3}. Section \uppercase\expandafter{\romannumeral4} formulates the problem, whereas Section \uppercase\expandafter{\romannumeral5} presents the contextual bandit-based solution strategy. Results are discussed and evaluated in Section \uppercase\expandafter{\romannumeral6}, followed by the conclusions.

\textit{Notation:} The variables in the day-ahead problem have the subscript $D$, while the look-ahead variables in the day-ahead problem have the subscript $\xi,D$. And the variables in the real-time problem have the subscript $R$.

\section{Motivating Examples}
Under the current operation framework of power systems, forecasts often serve as parameters in the subsequent decision-making problems. Thus, forecasting is not just related to how well the stochastic process of random variables is described. Instead, forecasting acquires its own value through the ability to influence the decisions made by the users of the forecast.

In this section, we present the point forecast for the UC problem and PI forecast for the wind power offering in the day-ahead market to show the necessity of value-oriented forecast.

\textbf{Example 1 (The point forecast of net load for UC problem):} Here, we consider a one-bus system with two generators (G1 and G2) serving the net load. Let
$x_1,x_2$ and $c_1,c_2$ denote the power generation and cost coefficients of the two generators, respectively. And let $u_1,u_2$ be the binary variables regarding the on/off status of the two generators and $d_1,d_2$ be the startup costs. Given the point forecast of the net load $\hat{l}$, the UC problem is formulated as,

\begin{subequations}\label{1}
\begin{align} &\mathop{\min}_{x_1,x_2,u_1,u_2}c_1\cdot x_1+c_2\cdot x_2+d_1\cdot u_1+d_2\cdot u_2\label{1(a)}\\ 
    &s.t. 0\leq x_1\leq \overline{x}_1\cdot u_1\label{1(b)}
    \\ 
    &\quad \ 0\leq x_2\leq \overline{x}_2\cdot u_2\label{1(c)}\\
    &\quad \ x_1+x_2=\hat{l}\label{1(d)},
\end{align}
\end{subequations}
where $\overline{x}_1,\overline{x}_2$ are the generation limits of the two generators. Here we assume G1 is cheaper than G2. And we assign the cost coefficients $c_1,c_2,d_1,d_2$ with the values 0.1 \$/kW, 0.2 \$/kW, 10 \$, 20 \$, and the generation limits $\overline{x}_1,\overline{x}_2$ with the values 50 kW, 40 kW.

Let the realization of net load $l$ be 49 kW. If the forecast of load $\hat{l}$ is 47 kW, the optimal solution of the UC problem is $x_1^*=47,x_2^*=0,u_1^*=1,u_2^*=0$. And once the realization of the net load $l$ is available, G1 generates an additional 2 kW of electricity to satisfy the demand. Therefore, the total cost under the forecast $\hat{l}=47$ kW is 14.9 \$. If the forecast of load $\hat{l}$ is 51 kW, the optimal solution of the UC problem is $x_1^*=50,x_2^*=1,u_1^*=1,u_2^*=1$. And once the realization of the net load $l$ is available, G1 and G2 respectively reduce 1 kW of electricity generation to satisfy the demand. And the total cost under the forecast $\hat{l}=51$ kW is 34.9 \$. It can be observed that although the two forecasts have the same deviation from the realization and therefore they have the same quality evaluated by the statistical quality metric such as mean squared error, the costs incurred by them are different. Therefore, this case shows the value-oriented forecast is important to the UC problem and the similar opinion can be found in \cite{chen2022feature}.

\textbf{Example 2 (PI forecast for wind power offering in day-ahead market):} Let $\hat{E}$ denote the quantity the wind power producer offers in the day-ahead market and $E$ denote the wind power realization. Under the day-ahead electricity price $\lambda^D$, the profit obtained in the day-ahead market is $\lambda^D\cdot\hat{E}$. In the two-price balance market where $\lambda^{UP},\lambda^{DW}$ are the prices for up- and down- regulation, the wind power producer has to buy up-regulation power when its actual realization $E$ is smaller than the offer $\hat{E}$, while down-regulation is to be sold when $E$ is larger than $\hat{E}$. The total profit of the wind power producer in the day-ahead and balance markets is therefore formulated as,
\begin{equation}\label{2}
\rho=\lambda^D\cdot\hat{E}+\lambda^{UP}\cdot[E-\hat{E}]^-+\lambda^{DW}\cdot[E-\hat{E}]^+,
\end{equation}
where $[ \cdot ]^-=\mathop{\min}(\cdot,0)$ and $[ \cdot ]^+=\mathop{\max}(\cdot,0)$. Eq. \eqref{2} can be equivalently formulated as,
\begin{equation}\label{3}
\rho=\lambda^D\cdot E-[(\lambda^D-\lambda^{UP})\cdot[E-\hat{E}]^-+(\lambda^D-\lambda^{DW})\cdot[E-\hat{E}]^+],
\end{equation}

Pricing rules entail that $\lambda^D \leq \lambda^{UP}$, $\lambda^D \geq \lambda^{DW}$. Given $[E-\hat{E}]^- \leq 0$ and $[E-\hat{E}]^+ \geq 0$, both terms inside the brackets are nonnegative. $\lambda^{UP}-\lambda^D$ and $\lambda^D-\lambda^{DW}$ are the costs of opportunity loss per energy unit under up-regulation and down-regulation, respectively. The wind power producer aims to offer the wind power $\hat{E}$ for maximizing the profit $\rho$.

Let us assume the values of the day-ahead electricity price $\lambda^D$, the prices of up- and down- regulation $\lambda^{UP},\lambda^{DW}$, and the wind power realization are 60 \$/MW, 300 \$/MW, 10 \$/MW, and 20 MW respectively. Denote the lower and upper bounds of PI for quantifying the wind power $E$ in day-ahead market as $\underline{q},\overline{q}$. The following PI-based robust optimization problem to determine the wind power offer in the worst case is formulated as,

\begin{equation}\label{4}
\begin{split}
\mathop{\max}_{\hat{E}\in[\underline{q},\overline{q}]}\mathop{\min}_{E\in[\underline{q},\overline{q}]}\lambda^D\cdot E-[(\lambda^D-\lambda^{UP})\cdot[E-\hat{E}]^-+\\
(\lambda^D-\lambda^{DW})\cdot[E-\hat{E}]^+]  
\end{split}
\end{equation}

Here, we consider the two day-ahead PI forecasts for the wind power $E$, namely [16,18] and [21,22]. Obviously, the statistical quality of the latter is better than the former. 

For the first PI forecast [16,18], in the worst scenario, the wind power offer equals 16 MW, and the optimal objective of \eqref{3} equals 960 \$. Then, when the wind power production is revealed in the real-time market, the profit incurred by down-regulation is 40 \$. Therefore, the total profit of the wind power producer in the day-ahead and real-time markets is 1000 \$.

For the second PI forecast [21,22], in the worst scenario, the wind power offer equals 21 MW, and the optimal objective of \eqref{4} equals 1260 \$. Then, when the wind power production is revealed in the real-time market, the profit incurred by down-regulation is -300 \$. Therefore, the total profit of the wind producer in the day-ahead and real-time markets is 960 \$, which is lower than the profit obtained under the first PI forecast. 

Therefore, the example shows that the good statistical quality cannot ensure the good value for the decision-making. As such, the value-oriented PI forecasting is needed to bridge the gap between the forecast and the decision.

\section{Preliminaries}

In this section, first we introduce the formulation of PI and discuss the need of value-oriented PI in subsection \emph {A}. In subsection \emph {B}, we introduce the two-timescale VPP operation, which is the downstream decision-making problem of the PI estimation task.

\subsection{Preliminary of Prediction Interval}
Let $Y_{t+k}$ denote a random variable for the target wind power output at future time $t+k$, $F_{Y_{t+k}}$ be the corresponding cumulative distribution function, and $y_{t+k}$ be the realization. Specifically, a PI with NCP $(1-\beta)\times 100 \%$ provides a summary of the cumulative distribution function $F_{Y_{t+k}}$, and can be developed as,
\begin{equation}\label{5} 
\begin{split}
&\mathcal{Y}_{t+k}=[\hat{q}_{t+k}^{\underline{\alpha}_{t+k}},\hat{q}_{t+k}^{\overline{\alpha}_{t+k}}]\\
&\overline{\alpha}_{t+k}=\underline{\alpha}_{t+k}+1-\beta,
\end{split}
\end{equation}
where $\underline{\alpha}_{t+k}$ is in the range of $(0,\beta)$, and $\hat{q}_{t+k}^{\underline{\alpha}_{t+k}}$, $\hat{q}_{t+k}^{\overline{\alpha}_{t+k}}$ are the predictions of the quantiles $F_{Y_{t+k}}^{-1}(\underline{\alpha}_{t+k})$, $F_{Y_{t+k}}^{-1}(\overline{\alpha}_{t+k})$. Given the probability proportion $\alpha \in \{\underline{\alpha}_{t+k},\overline{\alpha}_{t+k}\}$ and contextual information $\bm{s}_t$ up to time $t$, the quantile prediction can be achieved by training a QR model $f^{\alpha}(\bm{s}_t;\Theta^{\alpha})$ minimizing the pinball loss function, where QR models can be chosen as many off-the-shelf ones. It is described as,

\begin{equation}\label{6}    \hat{\Theta}^{\alpha}=\mathop{\arg\min}_{\Theta^{\alpha}}\mathbb{E}_{F_{Y_{t+k}}}[\ell^{\alpha}(f^{\alpha}(\bm{s}_t;\Theta^{\alpha}),Y_{t+k})],
\end{equation}
where $\ell^{\alpha}$ is the pinball loss function, defined as,

\begin{equation}\label{7}    \ell^{\alpha}(x,y)=\mathop{\max}\{\alpha(y-x),(\alpha-1)(y-x)\}.
\end{equation}

After the model training process illustrated in \eqref{6}, with the estimated model parameters $\hat{\Theta}^{\underline{\alpha}_{t+k}}$ and $\hat{\Theta}^{\overline{\alpha}_{t+k}}$, the predicted quantiles are given by $\hat{q}_{t+k}^{\underline{\alpha}_{t+k}}=f^{\underline{\alpha}_{t+k}}(\bm{s}_t;\hat{\Theta}^{\underline{\alpha}_{t+k}})$ and $\hat{q}_{t+k}^{\overline{\alpha}_{t+k}}=f^{\overline{\alpha}_{t+k}}(\bm{s}_t;\hat{\Theta}^{\overline{\alpha}_{t+k}})$.

The probability proportion $\underline{\alpha}_{t+k}$ is often chosen as $\beta/2$ if the distribution function $F_{Y_{t+k}}$ is symmetric, or chosen optimally and adaptively to the skewed distribution. Indeed, PI serves as an input to the subsequent decision-making. The optimality to the probability distribution cannot always ensure the optimality to the value of the downstream decision task. To tackle this challenge, we seek to find the optimal probability proportion for PI prediction, such that the value of the downstream decision task is maximized. Before we show how to achieve this in the context of two-timescale VPP operation, we firstly introduce the general model of its operation framework under the uncertainty of wind power output in the next subsection.

\begin{figure}[h]
  \centering
\includegraphics[scale=0.51]{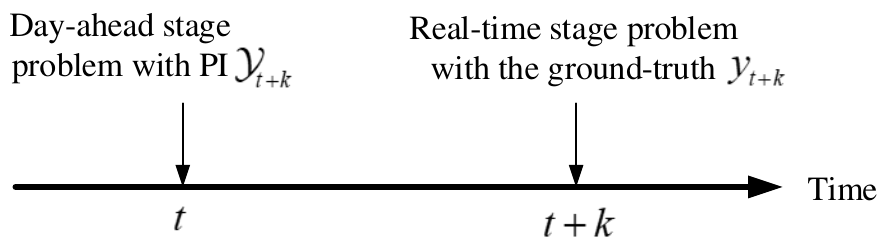}\\
\caption{Illustration of two-timescale operation.}\label{Fig 1}
\end{figure}

\subsection{Two-timescale VPP Operation under Wind Power Uncertainty}

In this work, we focus on issuing value-oriented PI for a two-timescale VPP operation task. The illustration of the day-ahead and real-time decision-making processes is shown Fig. \ref{Fig 1}, where the day-ahead and real-time decisions are made at different time. The day-ahead decision $\bm{x}_D$ is made to determine the generation levels and market bidding quantity under the uncertain wind power output $Y_{t+k}$ to satisfy the load demand $l_{t+k}$. Likewise, the decision $p_D$ regarding the day-ahead schedule of wind power is also made in the day-ahead problem. Here, we assume the load forecast is rather accurate and the future wind power output is the only source of uncertainty. The real-time stage decision $\bm{z}_R$ is made after knowing the wind power realization $y_{t+k}$, which settles the forecast deviation. As the decision $p_D$ is related with the forecast deviation, the day-ahead decision has impact on the decision of the real-time problem. To make the day-ahead decision optimal, the relationship among the day-ahead and real-time decision variables should be properly accounted for by solving a single optimization problem at a day-ahead time. In this sense, the real-time stage in the single optimization problem is a look-ahead one, which is referred to as a recourse problem in the literature \cite{morales2013integrating}. By describing the random variable $Y_{t+k}$ using an uncertainty set which is the prediction interval $\mathcal{Y}_{t+k}$ in this work, the day-ahead problem is a robust optimization with recourse. The general form is given by,

\begin{subequations}\label{8}
\begin{align} &\mathop{\min}_{\bm{x}_D,p_D}f_0^D(\bm{x}_D)+\mathcal{Q}_{\xi}^D\label{8(a)}\\ 
    &s.t. f_i^D(\bm{x}_D)\leq 0,\forall i \in I^D \label{8(b)}
    \\ 
    &\quad \ g^D(p_D)\leq 0\label{8(c)}\\
    &\quad \ h^D(\bm{x}_D)+p_D=l_{t+k}\label{8(d)},
\end{align}
\end{subequations}
where $I^D$ is the set of day-ahead inequality constraints regarding $\bm{x}_D$, and $p_D$ is the day-ahead variable of scheduled wind power. Eqs \eqref{8(b)} and \eqref{8(c)} are the inequality constraints of capacity limits, and \eqref{8(d)} is the equality constraint ensuring power balance.  $\mathcal{Q}_{\xi}^D$ is the recourse problem for future time. Let $\xi$ denote the uncertain parameter, whose realization is within the uncertainty set $\mathcal{Y}_{t+k}$. The problem is described as,

\begin{subequations}\label{9}
\begin{align} &\mathcal{Q}_{\xi}^D:=\mathop{\max}_{\xi \in \mathcal{Y}_{t+k}}\mathop{\min}_{\bm{z}_{\xi,D}}f_0^R(\bm{z}_{\xi,D})\\ 
    &\qquad \qquad \quad s.t. f_i^R(\bm{z}_{\xi,D})\leq 0,\forall i \in I^R \label{9(b)}\\ 
    &\qquad \qquad \quad \quad \ h^R(\bm{z}_{\xi,D})+\xi-p_D=0,\label{9(c)}
\end{align}
\end{subequations}
where $\bm{z}_{\xi,D}$ is the recourse decision to be made, which is related with $\xi$. The objective of the day-ahead problem in \eqref{8(a)} is to make the best decisions represented by variables $\bm{x}_D,p_D$  for the worst realization of parameters $\xi$ in the PI $\mathcal{Y}_{t+k}$ and considering the recourse decision $\bm{z}_{\xi,D}$. Solving \eqref{8} produces the optimal day-ahead solution $\bm{x}_D^*,p_D^*$, with which the day-ahead operational cost is $f_0^D(\bm{x}_D^*)$. Parameterized by the optimal day-ahead wind power schedule $p_D^*$ and the wind power realization $y_{t+k}$, the real-time problem is, 
\begin{subequations}\label{10}
\begin{align} &\mathop{\min}_{\bm{z}_{R}}f_0^R(\bm{z}_{R})\\ 
    &s.t. f_i^R(\bm{z}_{R})\leq 0,\forall i \in I^R \label{10(b)}\\ 
    &\quad \ h^R(\bm{z}_{R})+y_{t+k}-p_D^*=0.\label{10(c)}
\end{align}
\end{subequations}

With the optimal real-time solution $\bm{z}_R^*$, the optimal objective of the real-time problem is $f_0^R(\bm{z}_R^*)$. Intuitively, the monetary score $s(y_{t+k},\mathcal{Y}_{t+k})$, which is the sum of day-ahead and real-time optimal objectives, can measure the value of PI $\mathcal{Y}_{t+k}$ to the two-timescale VPP operation.

\begin{equation}\label{11}    s(y_{t+k},\mathcal{Y}_{t+k})=f_0^D(\bm{x}_D^*)+f_0^R(\bm{z}_R^*),
\end{equation}
which is a negatively oriented score, i.e., the lower the score, the better the value to the two-timescale operation. Directly using the score in \eqref{11} as the loss function to train the model for issuing PI is difficult. Firstly, it is non-differentiable with respect to the PI $\mathcal{Y}_{t+k}$, and therefore the commonly used neural network-based models cannot be applied. Furthermore, besides the model parameters $\Theta^\alpha,\forall \alpha \in \{\overline{\alpha}_{t+k},\underline{\alpha}_{t+k}\}$, it involves other decision variables regarding the system operation. As the models with parameters $\Theta^\alpha$ are commonly non-convex, simultaneously optimizing those variables is challenging. 

Certainly, the decision-making problem is not limited to the one needing wind power forecast. Indeed, the decision problem involving other random variables such as the solar power or the net load can be applied as well, where the parameter to be predicted lies in the right-hand side of the constraint  and needs a recourse, i.e., remedial actions once the values of the random variables become known and the different types of remedial actions result in different costs.

Remark 1: Although the adopted QR model has single output, it is also suitable for the task of multi-horizon forecasting. For instance, for the day-ahead forecasting of wind power where one wishes to obtain the forecasts of 24 hours, each forecast can be obtained point by point, once the contextual information for each target is available. In this line, the model, such as recurrent neural networks or Transformers, can also be applied and has the network structure of multiple inputs and a single output. Furthermore, as the decision-making problem in our study does not involve temporally correlated constraints, the QR model with a single output is suitable for the application. For the decision-making problem with temporally correlated constraints, the forecast model with multiple outputs may be needed for taking the temporal correlation into account.


\section{Methodological Framework}
In this work, we propose to use the contextual bandit-based approach to cope with the aforementioned challenges. Specifically, the negative value of the score in \eqref{11} is used as the reward for guiding the proportion selection, rather than acting as the loss function directly. Therefore, there is no need to differentiate over the score. Also, in the contextual bandit framework, the model parameters $\Theta^\alpha$ and the variables regarding the system operation are solved in two separate but related models, which tackles the second challenge.

Let us focus on the training stage and formulate the model estimation problem of value-oriented PI. Concretely, it involves a policy learning model with the parameters $\bm{W}$ and QR models with the parameters $\Theta^\alpha, \forall \alpha \in \{\underline{\alpha}_{t+k},\overline{\alpha}_{t+k}\}$. With the samples on the training set $\{\bm{s}_{t},y_{t+k}\}_{t \in \mathcal{T}^{tr}}$, the policy learning model maps the contextual information $\bm{s}_t$ to a optimal proportion $\underline{\alpha}_{t+k}$. With the selected probability proportion, the corresponding QR models are specified and output the predicted quantiles to form the LUB of PI $\mathcal{Y}_{t+k}$. The parameters of the QR models are learned by minimizing the pinball loss defined in \eqref{6}. And the parameters $\bm{W}$ of the policy learning model is optimized to minimize the sum of the score $s(y_{t+k},\mathcal{Y}_{t+k})$ in the training set. The calculation of the score relies on the predicted PI $\mathcal{Y}_{t+k}$ and the ground-truth label $y_{t+k}$ as inputs. To achieve this goal, we propose to estimate those parameters in an iterative manner by leveraging a value-oriented contextual bandit. And we detail the solution strategy in the next section.

Once the parameters of the policy learning model and QR models are obtained, i.e., $\hat{\bm{W}},\hat{\Theta}^\alpha, \forall \alpha \in \{\underline{\alpha}_{t+k},\overline{\alpha}_{t+k}\}$, in the operational forecasting stage, the policy learning function with the estimated parameters $\hat{\bm{W}}$ firstly maps the contextual information $\bm{s}_t$ to the selected proportion $\underline{\alpha}_{t+k}$ for any sample on the test set $t \in \mathcal{T}^{te}$. Then, the quantiles forming the LUB of PI are issued by $\hat{q}_{t+k}^{\underline{\alpha}_{t+k}}=f^{\underline{\alpha}_{t+k}}(\bm{s}_t;\hat{\Theta}^{\underline{\alpha}_{t+k}})$ and $\hat{q}_{t+k}^{\overline{\alpha}_{t+k}}=f^{\overline{\alpha}_{t+k}}(\bm{s}_t;\hat{\Theta}^{\overline{\alpha}_{t+k}})$. With the predicted PI, the VPP operator then solves the day-ahead problem in \eqref{8} and the subsequent real-time problem in \eqref{10}.

\section{Solution Strategy}
In this section, we develop the estimation approach based on the value-based contextual bandit. The overall solution framework is presented in subsection \textit{A}. And the proposed algorithm is given in subsection \textit{B}.

\subsection{The Solution Framework}

Indeed, the parameters estimation problem formulated in the section \uppercase\expandafter{\romannumeral4} has three interrelated tasks, namely the task of learning the proportion selection policy, the determination of optimal solutions of the two-timescale operation, and the QR model estimation task. In this sense, we specify the three tasks to the corresponding elements of the contextual bandit framework and link them by the closed-loop feedback. Concretely, an agent, modelled by NN, is responsible for learning the proportion selection policy. With the selected proportion $\underline{\alpha}_{t+k}$ of the lower bound quantile, the corresponding QR models $f^{\underline{\alpha}_{t+k}},f^{\overline{\alpha}_{t+k}}$ in the environment output the quantiles $\hat{q}_{t+k}^{\underline{\alpha}_{t+k}}=f^{\underline{\alpha}_{t+k}}(\bm{s}_t;\hat{\Theta}^{\underline{\alpha}_{t+k}})$ and $\hat{q}_{t+k}^{\overline{\alpha}_{t+k}}=f^{\overline{\alpha}_{t+k}}(\bm{s}_t;\hat{\Theta}^{\overline{\alpha}_{t+k}})$ which form the LUB of PI $\mathcal{Y}_{t+k}$, and then the models are updated with gradient descent by minimizing the pinball loss. With the PI $\mathcal{Y}_{t+k}$ and the label $y_{t+k}$, the value of the PI is calculated by solving the score $s(y_{t+k},\mathcal{Y}_{t+k})$ defined in \eqref{11} based on the optimal solutions of the two-timescale operation. Then, the value of PI is used as the feedback reward to link the agent and the environment. The whole framework is illustrated in Fig. \ref{Fig 2}, and the key elements are defined as follows:

\begin{figure}[h]
  \centering
  \includegraphics[scale=0.38]{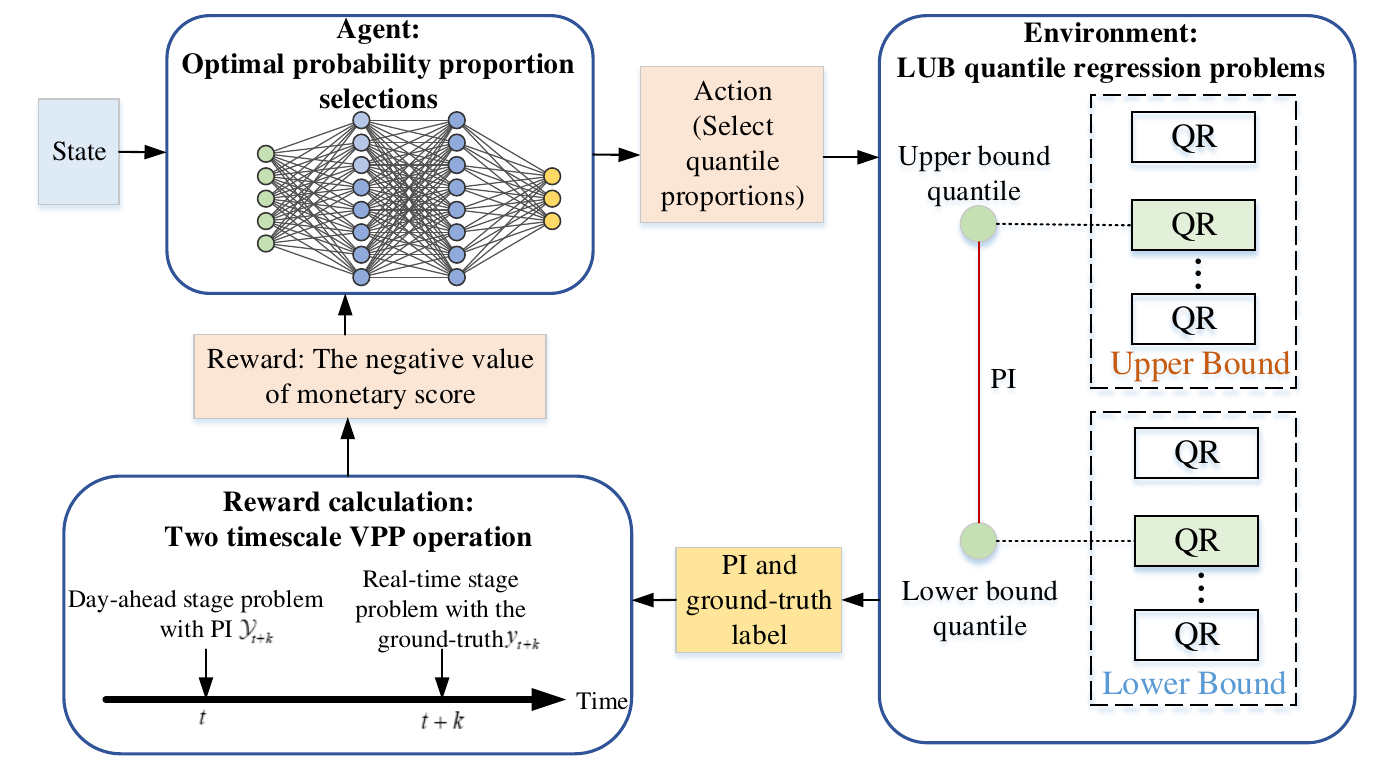}\\
  \caption{Illustration of contextual bandit-based value-oriented PI framework.}\label{Fig 2}
\end{figure}

\textit{State/Context}: 
The agent's input state $\bm{s}_t$ is a 4-dimensional vector, which is also the input feature of the QR model and is composed of the wind speed and direction at 10m and 100m altitude.

\textit{Action}: The continuous range of $(0,\beta)$ is discretized into several probability proportions, which form the action space $\mathcal{A}$ with the size $|\mathcal{A}|$. Given the certain NCP, the action space is \cite{zhang2022optimal},

\begin{equation}\label{12}
\left\{\frac{i\cdot\beta}{|\mathcal{A}|+1}\right\}^{|\mathcal{A}|}_{i=1},
\end{equation}
where $|\mathcal{A}|=2^n-1, n=1,2,...$ is the number of actions. The optimal proportion $\underline{\alpha}_{t+k}, \forall t \in \mathcal{T}^{tr}$ of the lower bound quantile is chosen from the set defined in \eqref{12} for each sample in the training set. And we denote $\bm{\underline{\alpha}}$ to be the vector   whose elements are the actions in the set $\mathcal{A}$,

\textit{Reward}: The monetary score defined in \eqref{11} is an evaluation metric of the value of PI to the two-timescale operation. As it is a negatively oriented score, the reward $r_t$ of the sample $t\in \mathcal{T}^{tr}$ in the training set is defined as its negative value,
\begin{equation}\label{13}
r_t = -s(y_{t+k},\mathcal{Y}_{t+k}).
\end{equation}

\textit{Environment}: Each probability proportion in the set $\mathcal{A}$ has a corresponding QR model. Therefore, there are $2|\mathcal{A}|$ QR models with its own parameters and buffer (which is a space to store the samples) for estimating the quantiles of LUB. And the task of model parameters estimation of the selected QR models is performed in the environment.
Since QR models for different proportions are trained separately, the quantile crossing effect may happen. However, the quantile crossing effect can be avoided under the proposed framework, by replacing the QR models in the environment with a model for the forecasting of continuous probability density, and the density model returns the quantile given the proportion determined by the agent.

\textit{Agent}: The optimal proportion selection policy is learned by an agent, which learns a mapping from the state $\bm{s}_t$ to the optimal proportion $\underline{\alpha}_{t+k}$ of the lower bound quantile.

Although the decision-making problem in our case study does not consider the constraints with temporal dependency, such a problem is also compatible with the proposed framework, once the policy learning algorithm of the agent is improved to adapt to the sparse reward. Let us consider a decision problem with temporal dependency on the time trajectory $\mathcal{T}$. For each time-step $t \in \mathcal{T}$, the agent chooses an action for specifying the proportion of the lower bound quantile. Then the quantiles of the lower and upper bounds are issued by the corresponding quantile regression models. In this line, the agent chooses $|\mathcal{T}|$ actions until receiving a reward at the end of the trajectory $\mathcal{T}$, which results in the sparse reward and makes the agent’s policy learning process difficult. Therefore, the policy learning algorithm coping with the sparse reward is needed to be investigated for coping with the decision-making problem with temporal dependency. Also, the contextual bandit, which is a special case of reinforcement learning focusing on the myopic consequence, is adopted here, since we focus on the decision problem with single time step. For the problem with temporally dependent constraints and multiple time steps, the reinforcement learning (RL) algorithm should be applied.

\subsection{Value-oriented PI Forecast with Value-based Contextual Bandit}

In this subsection, we first detail the value-based contextual bandit used in this work, and then propose the algorithm of value-oriented PI estimation based on it. 

The goal of value-based contextual bandit is to find the best policy under the contextual information that maximizes the total rewards in the training set, which corresponds to minimizing the sum of the score defined in \eqref{11}. And the agent learns a state-action approximation function $Q(\bm{s}_t,\underline{\alpha}_{t+k};\bm{W})$ to approximate the value of the reward $r_t$. In this sense, the agent performs the minimization of the estimation error,

\begin{equation}\label{14}
L=\mathop{\min}_{\bm{W}}\sum_{t \in \mathcal{T}^{tr}} \frac{1}{2}(Q(\bm{s}_t,\underline{\alpha}_{t+k};\bm{W})-r_t)^2,
\end{equation}
where the model $Q(\bm{s}_t,\underline{\alpha}_{t+k};\bm{W})$ is a NN. With the loss function defined in \eqref{14}, the batch optimization is used and the estimated parameters $\hat{\bm{W}}$ are updated by gradient descent based on a batch of data with size $B$ denoted by the set $B^Q=\{\bm{s}_t,r_t\}_{t=1}^B$, which is randomly sampled from the agent's buffer $D^Q$. 

\begin{equation}\label{15}   
\begin{split}
&\hat{\bm{W}} \leftarrow \\
&\hat{\bm{W}}-\eta_Q\cdot\sum_{t \in B^Q}(Q(\bm{s}_t,\underline{\alpha}_{t+k};\hat{\bm{W}})-r_t)\frac{\partial Q(\bm{s}_t,\underline{\alpha}_{t+k};\hat{\bm{W}})}{\partial \hat{\bm{W}}},
\end{split}
\end{equation}
where $\eta_Q$ is the learning rate. With the estimated parameters $\hat{\bm{W}}$, the agent chooses the action for the sample $t \in \mathcal{T}^{tr}$ in an $\epsilon$-greedy manner \cite{zhang2021closed}. That is, to balance the exploration and exploitation, there is $1-\epsilon$ probability to choose the optimal action $\underline{\alpha}_{t+k} = \mathop{\arg\max}Q(\bm{s}_t,\bm{\underline{\alpha}};\hat{\bm{W}})$ and there is $\epsilon$ probability to choose $\underline{\alpha}_{t+k}$ as a random one from the set $\mathcal{A}$. We note that with the estimated parameters $\hat{\bm{W}}$, the value of the actions in the set $\mathcal{A}$ under the state $\bm{s}_t$ can be estimated, and the selection policy is here implicit and can be derived directly from the value function in the $\epsilon$-greedy manner.

Once passing the action to the environment, the corresponding QR models output the predicted quantiles $\hat{q}_{t+k}^{\underline{\alpha}_{t+k}}=f^{\underline{\alpha}_{t+k}}(\bm{s}_t;\hat{\Theta}^{\underline{\alpha}_{t+k}}),\hat{q}_{t+k}^{\overline{\alpha}_{t+k}}=f^{\overline{\alpha}_{t+k}}(\bm{s}_t;\hat{\Theta}^{\overline{\alpha}_{t+k}})$ to form PI $\mathcal{Y}_{t+k}$. Then, the feature and label pair $(\bm{s}_t,y_{t+k})$ is stored in the buffers $D^{\underline{\alpha}_{t+k}},D^{\overline{\alpha}_{t+k}}$ of the QR models $f^{\underline{\alpha}_{t+k}}$,$f^{\overline{\alpha}_{t+k}}$. $\forall \alpha \in \{\underline{\alpha}_{t+k},\overline{\alpha}_{t+k}\}$, a batch of data $B^{\alpha}=\{\bm{s}_t,y_{t+k}\}_{t=1}^B$ are randomly sampled from the buffer $D^{\alpha}$ to update the selected QR model parameters by gradient descent, i.e., 

\begin{equation}\label{16}
\hat{\Theta}^\alpha \leftarrow \hat{\Theta}^\alpha-\eta_\alpha\cdot\sum_{t\in B^{\alpha}}\frac{\partial \ell^\alpha(f^\alpha(\bm{s}_t;\hat{\Theta}^\alpha),y_{t+k})}{\partial \hat{\Theta}^\alpha},
\end{equation}
where $\eta_\alpha$ is the learning rate. Using \eqref{16}, the parameters of the selected models are updated, while the parameters of the unselected ones in the environment remain the same. With the PI $\mathcal{Y}_{t+k}$ and the label $y_{t+k}$, the reward $r_t$ is calculated and stored with the state and action pair $(\bm{s}_t,\underline{\alpha}_{t+k},r_t)$ into the agent's buffer $D^Q$. The above process repeats until the algorithm converges. The pseudocode of the algorithm is summarized in Algorithm \ref{alg1}.

\begin{algorithm}[h]
	\caption{Value-oriented PI Estimation}
	\label{alg1}
	\begin{algorithmic}[1]
	\Require{Batch size $B$, learning rates $\eta_\alpha$,$\eta_Q$, NCP $1-\beta$, The number of epochs $E$}
	\State{Initialize $2\cdot|\mathcal{A}|$ QR models’ and agent’s parameters with random weights}
 
    \For{$e=1:E$}
    \For{$t=1:|\mathcal{T}^{tr}|$}
	    \State{Given the input state $\bm{s}_t$, with the probability $\epsilon$, the agent selects a random action; otherwise the agent selects $\underline{\alpha}_{t+k}=\mathop{\arg\max}Q(\bm{s}_t,\bm{\underline{\alpha}})$}
	    \Statex{\textit{// Update parameters for the selected QR models}}
	    \State{Execute the action in the environment: Select the QR models for LUB’s quantiles with probability proportions $\underline{\alpha}_{t+k}$ and $\overline{\alpha}_{t+k}$, and predict the corresponding quantiles: $\hat{q}_{t+k}^{\underline{\alpha}_{t+k}}=f^{\underline{\alpha}_{t+k}}(\bm{s}_t;\hat{\Theta}^{\underline{\alpha}_{t+k}}),\hat{q}_{t+k}^{\overline{\alpha}_{t+k}}=f^{\overline{\alpha}_{t+k}}(\bm{s}_t;\hat{\Theta}^{\overline{\alpha}_{t+k}})$}
	   \State{Respectively store the tuple $(\bm{s}_t,y_{t+k})$ in the selected predictors’ buffers $D^{\underline{\alpha}_{t+k}}$,$D^{\overline{\alpha}_{t+k}}$ }
	   \State{$\forall \alpha \in \{\underline{\alpha}_{t+k},\overline{\alpha}_{t+k}\}$, randomly sample a batch of data $B^\alpha$ from the buffer $D^\alpha$ and update the QR model parameters by \eqref{16}}
	   \Statex{\textit{// Update agent's parameters}}
	   \State{Calculate the reward in \eqref{13} and store the tuple $(\bm{s}_t,\underline{\alpha}_{t+k},r_t)$ in the agents buffer $D^Q$}
	   
	   \State{Sample random batch of data $B^Q$ from the buffer $D^Q$, and update the agent’s network using \eqref{15}}
	\EndFor 
    \EndFor
	\end{algorithmic}  
	
\end{algorithm}

\section{Case Study}

This section testifies the effectiveness of the proposed approach based on real-world datasets. The aim is to show the proposed approach (1) compared with quality-oriented PI methods, has larger value to the downstream operation task, (2) by integrating decision with forecasting, can achieve the lowest average monetary score under different NCP levels, and (3) has obvious operational advantage with large penetration of wind power. 

\subsection{Experimental Setups}
Before reporting results, we discuss the data, model details, benchmarks, and evaluation metrics in this subsection. Publicly available hourly wind data in the year of 2012 from GEFCom 2014 is utilized. In the day-ahead stage of two-timescale operation of VPP, the demand is satisfied by the wind power, the power output of two distributed generators (DGs), and the electricity bought from the day-ahead market. The yearly energy price data derived from a real electricity market is used here. The detailed two-timescale decision model and the corresponding parameters are given in Appendix \ref{Appendix A}.  

Since the specific model of agent and quantile predictor is not the main focus of the work, quantile multi-layer perception (QMLP) is used as the prediction model for the QR task, and the agent structure in dueling deep Q-network \cite{wang2016dueling} is applied here for the network structure of the agent in the value approximation task. Note that there is no restriction to the type of the NN-based model in the proposed framework, and many others can be applied here. The quantile predictors’ parameters are summarized in Appendix \ref{Appendix B}, along with the parameters of the agent, where the number of neurons in the output layer is equal to the number of actions. Here, we follow the common practice and set the hyperparameters to the default ones according to the past experiences of training QR models for the pinball loss minimization.

Five quality-oriented PI forecasting comparison candidates belonging to the categories of CPI and quality-oriented OPI are investigated. Concretely, the light quantile gradient boosting regression tree (Light QGBRT), QMLP guided by the pinball loss, and a naive benchmark \cite{wen2021probabilistic} that is an extension of the persistence model in probabilistic setting are the CPI comparison candidates. The quality-oriented OPI methods include the mixed integer programming model (MLMIP) \cite{zhao2020optimal} and the offline-trained reinforcement learning-based optimal and adaptive PI forecasting method (OAPI), whose agent has the same network structure as that in the proposed approach \cite{zhang2022optimal}. The evaluation of PI is from the two aspects on the test set, i.e., the quality and value. From the perspective of quality, we use the average coverage deviation (ACD), average width, and Winkler score to measure the calibration, sharpness, and overall skill of PI, respectively, where the negative ACD value implies the low reliability \cite{zhao2020optimal}, while for the average width and Winkler score, the lower, the better. From the perspective of value, we use the averaged value of the monetary score defined in \eqref{7} for evaluation, where the evaluation process is the same as the operational forecasting stage elaborated in section \uppercase\expandafter{\romannumeral4}.


The program is implemented on the laptop with Intel®CoreTM i5-10210U 1.6 GHz CPU, and 8.00 GM RAM\footnote{Codes will be available after publication.}.

\subsection{Operational Advantage of Value-oriented PI}

The PIs of wind power under 95$\%$ NCP are generated by the proposed approach and the five comparison candidates. Table \uppercase\expandafter{\romannumeral1} lists the quality and value metrics of PI, where the number of actions of the proposed approach and OAPI is set as three, and the capacity of wind power $P$ is scaled up to 30 MW. It is observed from Table \uppercase\expandafter{\romannumeral1} that evaluated from the perspective of overall quality, the QMLP approach produces the low-quality PI with the largest Winkler score, and the produced PIs have the largest average width. Compared with other CPI methods such as Light QGBRT, the results show the inferior model learning ability of the QMLP model. Since the CPI methods cannot adapt to the skewed statistical distribution, the CPI method based on Light QGBRT has larger Winkler score compared with the OAPI approach, although OAPI uses the QMLP model as its quantile predictor. The results show the importance of the adaptiveness to the skewness of the probability distribution, especially for the PI forecasting of wind power, whose distribution is generally positively skewed. Among the quality-oriented comparison candidates, the OAPI approach produces the best quality PI and the lowest monetary score. However, the relationship between the quality and value is complex, and merely improving the quality cannot ensure the improvement of value. Although the proposed value-oriented PI forecasting approach has the largest Winkler score and the negative ACD value, which indicates the poor quality, it has the lowest average monetary score among five comparison candidates. Therefore, the results display the operational advantage of value-oriented PI and call for the need to transit from quality-oriented PI forecast to value-oriented PI forecast. 

\begin{table}[h]
\caption{Forecasting quality and value of wind power PI under 95\% NCP}
\begin{center}
\begin{tabular}{c  c  c  c  c}
\hline\hline
    Method &
    \makecell[c]{Winkler \\score/MW} &
    \makecell[c]{Average \\width/MW} &
    ACD/\% &
    \makecell[c]{Average monetary\\ score/\$} \\
\hline
     The proposed & 32.33 &
     13.44 &
     -11 &
     1578 \\

     Light QGBRT & 23.54 &
     20.86 &
     1 &
     1590 \\

     QMLP & 29.15 &
     29.10 &
     4 &
     1602 \\

     Naive & 28.92 &
     28.72 &
     4 &
     1602 \\
     MLMIP & 27.47 &
     27.45 &
     4 &
     1601 \\
     OAPI & 23.38 &
     15.32 &
     -7 &
     1586 \\
\hline\hline
\end{tabular}
\end{center}
\end{table}

\begin{table}[h]
\caption{Training time of different PI forecasting models}
\begin{center}
\begin{tabular}{c  c  c  c  c}
\hline\hline
    \makecell[c]{The \\proposed} &
    \makecell[c]{Light \\QGBRT} &
    \makecell[c]{QMLP} &
    MLMIP &
    OAPI\\
\hline
     1h 10min &
     12s &
     1min &
     4min &
     51min\\
\hline\hline
\end{tabular}
\end{center}
\end{table}

The training time of different PI forecasting models is shown in Table \uppercase\expandafter{\romannumeral2}. As the naive method is a model-free one, the time is not listed here. The Light QGBRT method has the shortest training time, since it is a light model designed to be computationally efficient. Unsurprisingly, the proposed approach has the longest training time, as the training process involves estimating the model parameters and solving the decision problem in an iterative manner. However, the training time is acceptable for the offline training and online operational forecasting, and can be further shortened by using GPU.

\begin{figure}[h]
  \centering
\includegraphics[scale=0.7]{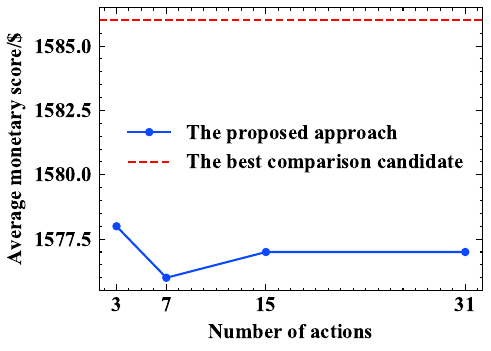}\\
\caption{Comparison of the operational value of the proposed approach under the different number of actions.}\label{Fig 3}
\end{figure}

Furthermore, the average monetary score of the proposed approach under different number of actions is shown in Fig. \ref{Fig 3}, where the red horizontal line displays the average monetary score of the best comparison candidate, which is the OAPI approach as shown in Table \uppercase\expandafter{\romannumeral1}. The average monetary scores of the proposed approach under different action space size are lower than the comparison candidate. Although there are some fluctuations, which is brought by the randomness of the $\epsilon$- greedy mechanism of the contextual bandit algorithm, the results under different action numbers are relatively stable. The results show that the number of actions is not a relevant hyperparameters impacting the performance. If the computational issue is a main concern, one may choose the number of QR models as three, which is the smallest number of actions.

The average monetary score can be decomposed into two parts, namely, the average of day-ahead monetary score $f_0^p(\bm{x}_t^*)$ and the average of real-time monetary score $f_0^o(\bm{z}_{t+k}^*)$. We show the decomposed average monetary score in Table \uppercase\expandafter{\romannumeral3}, where the negative real-time monetary score means that in real-time operation, the revenue brought by down-regulation is larger than the cost brought by up-regulation. We also show the result under deterministic wind power forecast in the last row, which has the largest average monetary score among all the methods. The results show the necessity of PI forecasting to capture the wind uncertainty, such that the uncertainty can be properly accounted for when making the day-ahead decision. Among the PI forecasting methods, due to  the forecasting process fully considering decision-making outcome, the proposed approach achieves the best operation value with the lowest average day-ahead monetary score among the PI-based methods.

\begin{table}[h]
\setlength{\tabcolsep}{1.5mm}{
\caption{Operational value of wind power PIs under different methods}
\begin{center}
\begin{tabular}{c  c  c  c}
\hline\hline
    Method &
    \makecell[c]{Average  \\monetary score/\$} &
    \makecell[c]{Average day-ahead \\monetary score/\$} &
    \makecell[c]{Average real-time\\ monetary score/\$} \\
\hline
     The proposed & 1578 &
     1682 &
     -104 \\

     Light QGBRT & 1590 &
     1733 &
     -143 \\

     QMLP & 1602 &
     1762 &
     -160  \\

     Naive & 1602 &
     1762 &
     -160 \\
     
     MLMIP & 1601 &
     1760 &
     -159 \\
     
     OAPI & 1586 &
     1721 &
     -135 \\
     Deterministic & 1606 &
     1474 &
     132 \\
\hline\hline
\end{tabular}
\end{center}}
\end{table}

\begin{figure}[h]
  \centering
\includegraphics[scale=0.46]{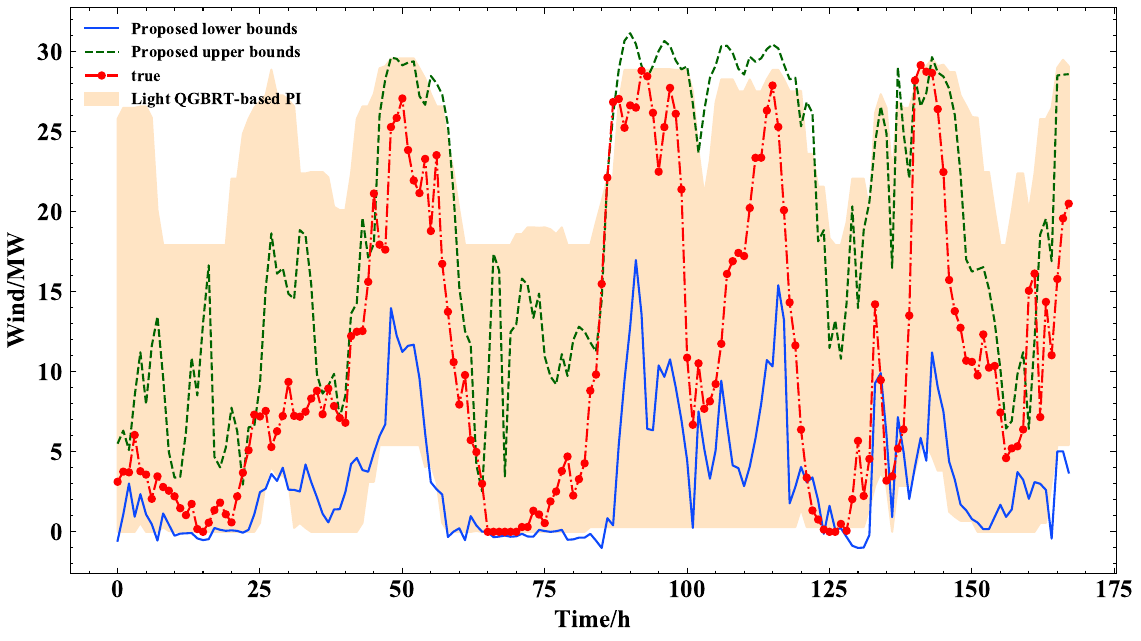}\\
  \caption{168 hours PIs forecast on test set obtained via the proposed approach and Light QGBRT.}\label{Fig 5}
\end{figure}

Fig. \ref{Fig 5} shows the 168 hours PIs on test set predicted by the proposed approach, superimposed with the PIs predicted by Light QGBRT. The LUB of the Light QGBRT generated PIs shows  insufficient adaptation to the change of wind power. As a result, the width of the PIs is relatively larger. In contrast, the proposed approach exhibits the time-varying PIs which track the change of the ground-truth, and therefore results in the smaller width.

To sum up, the proposed value-oriented PI has larger operational value to the VPP operation, and meanwhile ensures the excellent width performance of PIs.

\subsection{Investigation on Different Nominal Coverage Probability}

\begin{figure}[h]
  \centering
\includegraphics[scale=0.35]{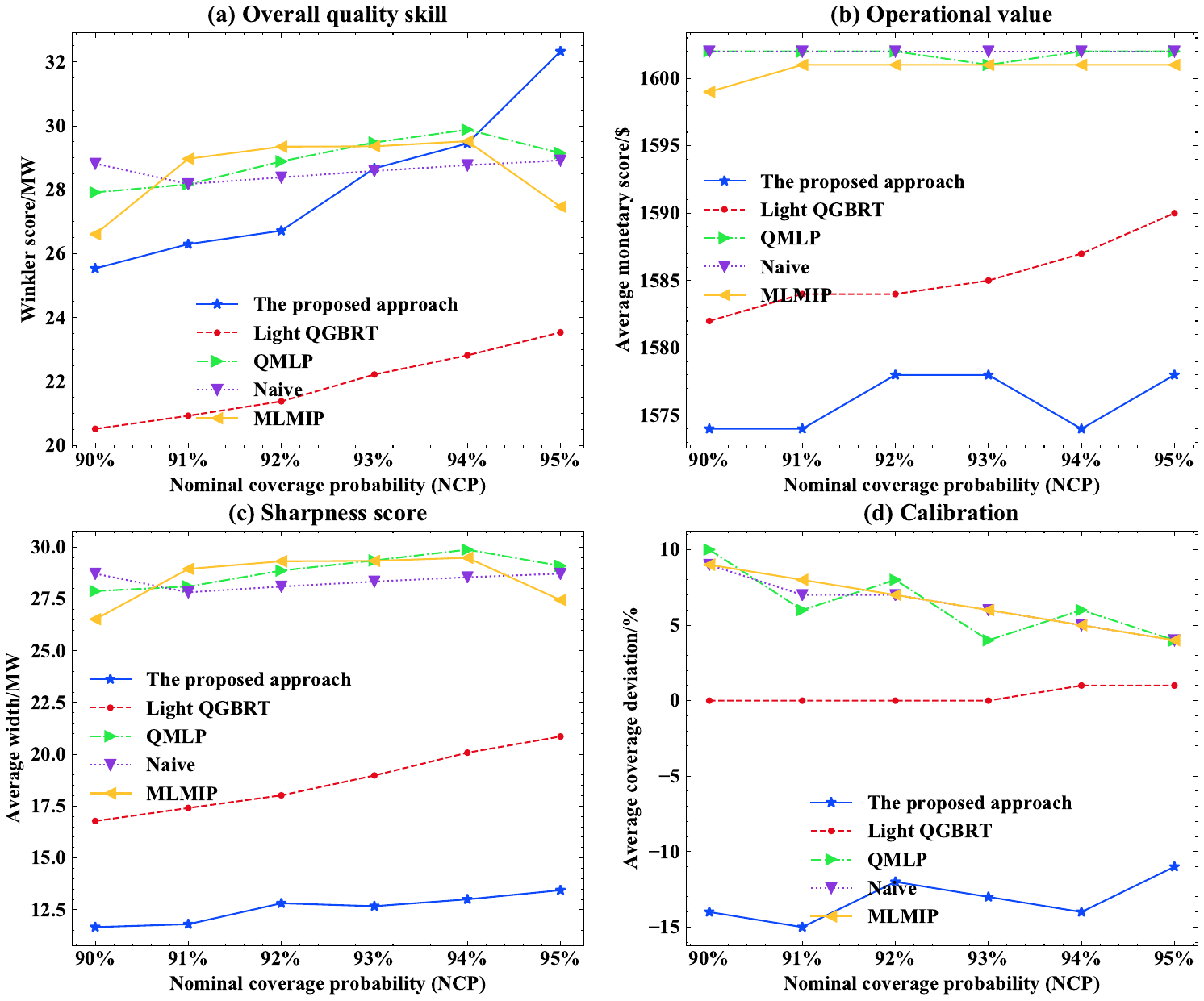}\\
  \caption{Forecasting quality and value of PIs under different NCPs.}\label{Fig 6}
\end{figure}

NCP has direct impact on the quality of PI. It is meaningful to investigate the value and quality of PIs under different NCP levels. Fig.\ref{Fig 6} shows the value and quality evaluation metrics of PIs under NCPs ranging from 90\% to 95\%. Those NCPs are frequently used in practice \cite{zhao2021cost}. In terms of the quality-oriented metrics, the average width of value-oriented PIs under large NCP is generally larger than the PIs under small NCP, as the larger NCP requires the higher probability that the realization falls into the range of PIs. For the metrics such as Winkler score and ACD whose calculation involves using NCP as a parameter, the value-oriented PIs have negative ACD score under different NCP levels, and the Winkler score increases with the increase of NCP.

The good quality of PIs cannot always be translated to better operational value, as the quality along with decision-making structure interactively influence the value of PIs to the operation. For instance, although the value-oriented PI forecasting have unsatisfying Winkler score under varying NCP levels, it always has the smallest average monetary score compared with other quality-oriented methods. Furthermore, there isn't a definite pattern between the operational value and NCP level. Therefore, it is necessary to reexamine the relationship between NCP and the value of PI, and design new value-oriented PI forecasting method based on it.

\subsection{Investigation on Different Wind Power Capacity}

The wind power capacity is important to two-timescale VPP operation and operational value evaluation. In this subsection, we investigate the operational value improvement of the proposed approach compared with quality-oriented candidates under different wind power capacity. Concretely, the wind power capacity is scaled up to 8MW and 30 MW respectively. And wind power PIs under 90\% and 95\% NCP are respectively employed in decision-making. The accumulative monetary score reduction on test set, which is the monetary score difference between the quality-oriented PI methods and the proposed approach, is shown in Tables \uppercase\expandafter{\romannumeral4} and \uppercase\expandafter{\romannumeral5}. The proposed approach shows the obvious accumulative monetary score reduction under wind power PIs with 90\% and 95\% NCP. In fact, due to considering the decision-making outcome into the forecast process, the proposed approach can capture the impact of the wind power capacity on the operational value. And the results show that the operational advantage of the proposed approach is more obvious under large wind power capacity. Therefore, with the increasing penetration of wind power, there is an urgent need to perform value-oriented PI forecast.

\begin{table}[h]
\caption{Accumulative monetary score reduction of wind power PI with 95\% NCP under different wind power capacity}
\begin{center}
\setlength{\tabcolsep}{1.5mm}{
\begin{tabular}{c  c  c  c c}
\hline\hline                
    \makecell[c]{Wind\\capacity} &
    \makecell[c]{Light QGBRT \\/The proposed} &
    \makecell[c]{QMLP \\/The proposed} &
    \makecell[c]{Naive \\/The proposed} &
    \makecell[c]{MLMIP \\/The proposed}\\
\hline
     8 MW & 7008 \$ &
     12264 \$&
     12264 \$&
     12264 \$\\
     
     30 MW & 21024 \$ &
     42048 \$&
     42048 \$&
     40296 \$ \\
\hline\hline
\end{tabular}}
\end{center}
\end{table}

\begin{table}[h]
\caption{Accumulative monetary score reduction of wind power PI with 90\% NCP under different wind power capacity}
\begin{center}
\setlength{\tabcolsep}{1.5mm}{
\begin{tabular}{c  c  c  c  c}
\hline\hline                
    \makecell[c]{Wind\\capacity} &
    \makecell[c]{Light QGBRT \\/The proposed} &
    \makecell[c]{QMLP \\/The proposed} &
    \makecell[c]{Naive \\/The proposed} &
    \makecell[c]{MLMIP \\/The proposed}\\
\hline
     8 MW & 7008 \$ &
     15768 \$&
     15768 \$&
     15768 \$\\
     
     30 MW & 14016	 \$ &
     49056 \$&
     49056 \$&
     43800 \$ \\
\hline\hline
\end{tabular}}
\end{center}
\end{table}

\subsection{Comparison with Other Contextual Bandit-based Approaches}
The contextual bandit algorithm used in this work is a special case of reinforcement learning (RL) without considering the look-ahead consequences. Therefore, the algorithm and network structure in RL can be applied here. In this subsection, we compare the current approach based on the dueling deep Q-network with the counterpart based on other RL algorithms. The traditional deep Q-network (DQN) and deep policy gradient (DPG) -based approaches are investigated here. Also, the result based on deep deterministic policy gradient (DDPG) with continuous action is obtained. The results on the averaged monetary score of PI with 95\% NCP are listed in Table \uppercase\expandafter{\romannumeral6}. 

\begin{table}[h]
\caption{Comparison on averaged monetary score of the value-oriented PI forecasts based on different RL approaches}
\begin{center}
\begin{tabular}{c  c  c c}
\hline\hline
The proposed & DQN & DPG & DDPG\\
\hline
1578 \$ & 1582 \$ & 1584 \$ & 1590 \$\\




\hline\hline
\end{tabular}
\end{center}
\end{table}

It is shown that the proposed approach has a lower monetary score compared with the other approaches, which indicates the superior performance in terms of the value. We observe that the DDPG-based approach has the largest score, and we infer that as the agent of the DDPG algorithm outputs continuous action in a range, it is more difficult for the agent to learn the policy compared with the policy learning of the agent outputting discrete action. The experiment also demonstrates that the agent outputting continuous action is compatible with the proposed framework, and more advanced RL algorithms are needed for improving the performance.

\section{Conclusion}

It is natural to consider the value of PI in the subsequent decision-making procedure, as good statistical scores cannot guarantee high value in the view of operation. In this paper, a value-oriented PI forecasting approach is proposed to fill the gap between forecasting and decision-making. It provides PIs with the aim of minimizing the operational cost. Particularly, the contextual bandit algorithm is leveraged to seek the best quantile proportions, where the agent learns the policy for probability proportion selection, and the corresponding QR is performed in environment. The two elements are linked by the negative value of the optimal decision-making objectives. 

Case study on a VPP operation problem with wind power reveals the occurrence of the quality/value “reversal”. That is, although the issued forecasts have the largest Winkler score, the operational cost of the subsequent decision-making is the smallest. Also, its good operational performance is demonstrated under different NCP levels. And the operational advantage of the proposed value-oriented PI approach is especially evident under larger penetration of wind power.

The goal of this paper is to call for an emphasis on linking forecasting and decision-making. It is necessary to develop other value-oriented forecasting products, such as point forecast, probability density forecast, to name a few, in the future work. Furthermore, the actions for selecting proper probability proportion in a range are discrete. Therefore, it is still required to explore the continuous action in the future. Besides, how to generate value-oriented PI for decision-making problems with temporal dependent constraints is also worth studying. And, although the main focus of this work is providing a value-oriented univariate PI forecasting,  the value-oriented multivariate PI forecasting is a challenging topic and deserves more investigation.

\section*{Acknowledgement}
The authors would like to appreciate the Shanghai Jiao Tong University Grants.

\bibliographystyle{IEEEtran}
\bibliography{IEEEabrv,mylib}

\begin{thebibliography}{10}
\providecommand{\url}[1]{#1}
\csname url@samestyle\endcsname
\providecommand{\newblock}{\relax}
\providecommand{\bibinfo}[2]{#2}
\providecommand{\BIBentrySTDinterwordspacing}{\spaceskip=0pt\relax}
\providecommand{\BIBentryALTinterwordstretchfactor}{4}
\providecommand{\BIBentryALTinterwordspacing}{\spaceskip=\fontdimen2\font plus
\BIBentryALTinterwordstretchfactor\fontdimen3\font minus
  \fontdimen4\font\relax}
\providecommand{\BIBforeignlanguage}[2]{{%
\expandafter\ifx\csname l@#1\endcsname\relax
\typeout{** WARNING: IEEEtran.bst: No hyphenation pattern has been}%
\typeout{** loaded for the language `#1'. Using the pattern for}%
\typeout{** the default language instead.}%
\else
\language=\csname l@#1\endcsname
\fi
#2}}
\providecommand{\BIBdecl}{\relax}
\BIBdecl

\bibitem{roald2022power}
L.~A. Roald, D.~Pozo, A.~Papavasiliou, D.~K. Molzahn, J.~Kazempour, and
  A.~Conejo, ``Power systems optimization under uncertainty: A review of
  methods and applications,'' in \emph{22nd Power Systems Computation
  Conference (PSCC)}, 2022.

\bibitem{8352041}
D.~Lee, H.~Shin, and R.~Baldick, ``Bivariate probabilistic wind power and
  real-time price forecasting and their applications to wind power bidding
  strategy development,'' \emph{IEEE Transactions on Power Systems}, vol.~33,
  no.~6, pp. 6087--6097, 2018.

\bibitem{2020Hong}
T.~Hong, P.~Pinson, Y.~Wang, R.~Weron, D.~Yang, and H.~Zareipour, ``Energy
  forecasting: A review and outlook,'' \emph{IEEE Open Access Journal of Power
  and Energy}, vol.~7, pp. 376--388, 2020.

\bibitem{8299478}
Y.~Wang, Q.~Chen, M.~Sun, C.~Kang, and Q.~Xia, ``An ensemble forecasting method
  for the aggregated load with subprofiles,'' \emph{IEEE Transactions on Smart
  Grid}, vol.~9, no.~4, pp. 3906--3908, 2018.

\bibitem{wen2022continuous}
H.~Wen, P.~Pinson, J.~Ma, J.~Gu, and Z.~Jin, ``Continuous and distribution-free
  probabilistic wind power forecasting: A conditional normalizing flow
  approach,'' \emph{IEEE Transactions on Sustainable Energy}, vol.~13, no.~4,
  pp. 2250--2263, 2022.

\bibitem{ZHANG2020105388}
Y.~Zhang, Q.~Ai, F.~Xiao, R.~Hao, and T.~Lu, ``Typical wind power scenario
  generation for multiple wind farms using conditional improved wasserstein
  generative adversarial network,'' \emph{International Journal of Electrical
  Power \& Energy Systems}, vol. 114, p. 105388, 2020.

\bibitem{8253871}
F.~Golestaneh, P.~Pinson, R.~Azizipanah-Abarghooee, and H.~B. Gooi,
  ``Ellipsoidal prediction regions for multivariate uncertainty
  characterization,'' \emph{IEEE Transactions on Power Systems}, vol.~33,
  no.~4, pp. 4519--4530, 2018.

\bibitem{5451173}
P.~Pinson and G.~Kariniotakis, ``Conditional prediction intervals of wind power
  generation,'' \emph{IEEE Transactions on Power Systems}, vol.~25, no.~4, pp.
  1845--1856, 2010.

\bibitem{zhao2021operating}
C.~Zhao, C.~Wan, and Y.~Song, ``Operating reserve quantification using
  prediction intervals of wind power: An integrated probabilistic forecasting
  and decision methodology,'' \emph{IEEE Transactions on Power Systems},
  vol.~36, no.~4, pp. 3701--3714, 2021.

\bibitem{zhao2021cost}
C.~Zhao, C.~wan, and Y.~Song, ``Cost-oriented prediction intervals: On bridging
  the gap between forecasting and decision,'' \emph{IEEE Transactions on Power
  Systems}, vol.~37, no.~4, pp. 3048--3062, 2022.

\bibitem{attarha2018adaptive}
A.~Attarha, N.~Amjady, S.~Dehghan, and B.~Vatani, ``Adaptive robust
  self-scheduling for a wind producer with compressed air energy storage,''
  \emph{IEEE Transactions on Sustainable Energy}, vol.~9, no.~4, pp.
  1659--1671, 2018.

\bibitem{bertsimas2012adaptive}
D.~Bertsimas, E.~Litvinov, X.~A. Sun, J.~Zhao, and T.~Zheng, ``Adaptive robust
  optimization for the security constrained unit commitment problem,''
  \emph{IEEE transactions on power systems}, vol.~28, no.~1, pp. 52--63, 2012.

\bibitem{qiu2018interval}
H.~Qiu, W.~Gu, Y.~Xu, Z.~Wu, S.~Zhou, and J.~Wang, ``Interval-partitioned
  uncertainty constrained robust dispatch for ac/dc hybrid microgrids with
  uncontrollable renewable generators,'' \emph{IEEE Transactions on Smart
  Grid}, vol.~10, no.~4, pp. 4603--4614, 2018.

\bibitem{wang2019probabilistic}
Y.~Wang, D.~Gan, M.~Sun, N.~Zhang, Z.~Lu, and C.~Kang, ``Probabilistic
  individual load forecasting using pinball loss guided lstm,'' \emph{Applied
  Energy}, vol. 235, pp. 10--20, 2019.

\bibitem{zhang2020admm}
Y.~Zhang, Q.~Ai, and Z.~Li, ``Admm-based distributed response quantity
  estimation: a probabilistic perspective,'' \emph{IET Generation, Transmission
  \& Distribution}, vol.~14, no.~26, pp. 6594--6602, 2020.

\bibitem{wen2021probabilistic}
H.~Wen, J.~Gu, J.~Ma, L.~Yuan, and Z.~Jin, ``Probabilistic load forecasting via
  neural basis expansion model based prediction intervals,'' \emph{IEEE
  Transactions on Smart Grid}, vol.~12, no.~4, pp. 3648--3660, 2021.

\bibitem{4530750}
H.~Bludszuweit, J.~A. Dominguez-Navarro, and A.~Llombart, ``Statistical
  analysis of wind power forecast error,'' \emph{IEEE Transactions on Power
  Systems}, vol.~23, no.~3, pp. 983--991, 2008.

\bibitem{zhao2020optimal}
C.~Zhao, C.~Wan, Y.~Song, and Z.~Cao, ``Optimal nonparametric prediction
  intervals of electricity load,'' \emph{IEEE Transactions on Power Systems},
  vol.~35, no.~3, pp. 2467--2470, 2020.

\bibitem{zhang2022optimal}
Y.~Zhang, H.~Wen, Q.~Wu, and Q.~Ai, ``Optimal adaptive prediction intervals for
  electricity load forecasting in distribution systems via reinforcement
  learning,'' \emph{arXiv preprint arXiv:2205.08698}, 2022.

\bibitem{pearce2018high}
T.~Pearce, A.~Brintrup, M.~Zaki, and A.~Neely, ``High-quality prediction
  intervals for deep learning: A distribution-free, ensembled approach,'' in
  \emph{International conference on machine learning}.\hskip 1em plus 0.5em
  minus 0.4em\relax PMLR, 2018, pp. 4075--4084.

\bibitem{wan2013optimal}
C.~Wan, Z.~Xu, P.~Pinson, Z.~Y. Dong, and K.~P. Wong, ``Optimal prediction
  intervals of wind power generation,'' \emph{IEEE Transactions on Power
  Systems}, vol.~29, no.~3, pp. 1166--1174, 2013.

\bibitem{murphy1993good}
A.~H. Murphy, ``What is a good forecast? an essay on the nature of goodness in
  weather forecasting,'' \emph{Weather and forecasting}, vol.~8, no.~2, pp.
  281--293, 1993.

\bibitem{morales2013integrating}
J.~M. Morales, A.~J. Conejo, H.~Madsen, P.~Pinson, and M.~Zugno,
  \emph{Integrating renewables in electricity markets: operational
  problems}.\hskip 1em plus 0.5em minus 0.4em\relax Springer Science \&
  Business Media, 2013, vol. 205.

\bibitem{9716858}
A.~C. Stratigakos, S.~Camal, A.~Michiorri, and G.~Kariniotakis, ``Prescriptive
  trees for integrated forecasting and optimization applied in trading of
  renewable energy,'' \emph{IEEE Transactions on Power Systems}, pp. 1--1,
  2022.

\bibitem{8706264}
T.~Carriere and G.~Kariniotakis, ``An integrated approach for value-oriented
  energy forecasting and data-driven decision-making application to renewable
  energy trading,'' \emph{IEEE Transactions on Smart Grid}, vol.~10, no.~6, pp.
  6933--6944, 2019.

\bibitem{chen2022feature}
X.~Chen, Y.~Yang, Y.~Liu, and L.~Wu, ``Feature-driven economic improvement for
  network-constrained unit commitment: A closed-loop predict-and-optimize
  framework,'' \emph{IEEE Transactions on Power Systems}, vol.~37, no.~4, pp.
  3104--3118, 2022.

\bibitem{pp}
P.~Pinson, ``Wind energy: Forecasting challenges for its operational
  management,'' \emph{Statistical Science}, vol.~28, no.~4, pp. 564 -- 585,
  2013.

\bibitem{NIPS2017_3fc2c60b}
P.~Donti, B.~Amos, and J.~Z. Kolter, ``Task-based end-to-end model learning in
  stochastic optimization,'' in \emph{Advances in Neural Information Processing
  Systems}, vol.~30.\hskip 1em plus 0.5em minus 0.4em\relax Curran Associates,
  Inc., 2017.

\bibitem{elmachtoub2022smart}
A.~N. Elmachtoub and P.~Grigas, ``Smart “predict, then optimize”,''
  \emph{Management Science}, vol.~68, no.~1, pp. 9--26, 2022.

\bibitem{zhang2022cost}
J.~Zhang, Y.~Wang, and G.~Hug, ``Cost-oriented load forecasting,''
  \emph{Electric Power Systems Research}, vol. 205, p. 107723, 2022.

\bibitem{bottieau2021automatic}
J.~Bottieau, K.~Bruninx, A.~Sanjab, Z.~De~Gr{\`e}ve, F.~Vall{\'e}e, and J.-F.
  Toubeau, ``Automatic risk adjustment for profit maximization in renewable
  dominated short-term electricity markets,'' \emph{International Transactions
  on Electrical Energy Systems}, vol.~31, no.~12, p. e13152, 2021.

\bibitem{zhang2021closed}
Y.~Zhang, Q.~Wu, Q.~Ai, and J.~P. Catal{\~a}o, ``Closed-loop aggregated
  baseline load estimation using contextual bandit with policy gradient,''
  \emph{IEEE Transactions on Smart Grid}, vol.~13, no.~1, pp. 243--254, 2021.

\bibitem{wang2016dueling}
Z.~Wang, T.~Schaul, M.~Hessel, H.~Hasselt, M.~Lanctot, and N.~Freitas,
  ``Dueling network architectures for deep reinforcement learning,'' in
  \emph{International conference on machine learning}.\hskip 1em plus 0.5em
  minus 0.4em\relax PMLR, 2016, pp. 1995--2003.

\end{thebibliography}

\appendices
\section{Descriptions of two-timescale VPP Operation Problem}\label{Appendix A}

The detailed form of the day-ahead robust optimization with recourse is given in \eqref{17}. Let $x_{1,D},x_{2,D}$ denote the power generation of DG1 and DG2 and $x_{3,D}$ denote the power purchased from the market under the price $\pi_{t+k}$, where $\bm{x}_D=[x_{1,D},x_{2,D},x_{3,D}]$. The output capacity is denoted by $\bar{x}_i,\forall i \in \{1,2\}$. The quadratic generation cost form part of the objective function in \eqref{17(a)}. 

\begin{subequations}\label{17}
\begin{align} &\mathop{\min}_{\bm{x}_D,p_D}\sum_{i=1}^2\frac{1}{2}a_i \cdot x_{i,D}^2+b_i \cdot x_{i,D}+c_i+x_{3,D}\cdot \pi_{t+k}+\mathcal{Q}_{\xi}^D\label{17(a)}\\ 
    &s.t. 0 \leq x_{i,D} \leq \bar{x}_i,\forall i \in \{1,2\} \label{17(b)}
    \\ 
    &\quad \ 0 \leq p_D \leq P\label{17(c)}\\
    &\quad \ x_{1,D}+x_{2,D}+x_{3,D}+p_D=l_{t+k}\label{17(d)},
\end{align}
\end{subequations}
where \eqref{17(b)} gives the constraint of maximum output power of DG, and \eqref{17(c)} limits that the scheduled day-ahead wind power should be less than the wind capacity $P$. And \eqref{17(d)} gives the power balance constraint. For the recourse problem, let $z_{i,\xi,D}^U,\forall i \in {1,2}$ denote the block of up-regulation power, and $z_{i,\xi,D}^D,\forall i \in {1,2}$ denote the block of down- regulation power, where $\bm{z}_{\xi,D}=[z_{1,\xi,D}^U,z_{2,\xi,D}^U,z_{1,\xi,D}^D,z_{2,\xi,D}^D]$. Let $c_{D,i},c_{U,i}$ denote the down- and up- unit regulation cost. With the objective of minimizing the operation cost, the look-ahead recourse problem $\mathcal{Q}_{\xi}^D$ is defined as,

\begin{subequations}\label{18}
\begin{align} &\mathop{\max}_{\xi\in \mathcal{Y}_{t+k}}\mathop{\min}_{\bm{z}_{\xi,D}}\sum_{i=1}^2-c_{D,i}z_{i,\xi,D}^D+\sum_{i=1}^2c_{U,i}z_{i,\xi,D}^U\\ 
    &\qquad \qquad \quad s.t. 0\leq z_{i,\xi,D}^D \leq \bar{z}_i^D,\forall i \in \{1,2\} \label{18(b)}\\ 
    &\qquad \qquad \quad \quad \ 0\leq z_{i,\xi,D}^U \leq \bar{z}_i^U,\forall i \in \{1,2\} \label{18(c)}\\
    &\nonumber \qquad \qquad \quad \quad \ -\sum_{i=1}^2 z_{i,\xi,D}^D + \sum_{i=1}^2 z_{i,\xi,D}^U +  \\
    & \qquad \qquad \qquad \qquad \qquad \xi-p_D = 0\label{18(d)},
\end{align}
\end{subequations}
where \eqref{18(b)} and \eqref{18(c)} are the constraints regarding the output limits of regulation blocks. Eq. \eqref{18(d)} ensures that the deviation is settled by the outputs of regulation blocks. 

The formulation of the real-time problem in \eqref{10} is similar to the inner minimization problem of \eqref{18} by replacing $\bm{z}_{\xi,D}$ with $\bm{z}_{R}$. And the values of parameters in \eqref{17} and \eqref{18} are shown in Table \uppercase\expandafter{\romannumeral7}, Table \uppercase\expandafter{\romannumeral8}, and Table \uppercase\expandafter{\romannumeral9}.

\begin{table}
\caption{Cost and technical data of DGs of VPP system}
\begin{center}
\renewcommand\arraystretch{1.5}{
\begin{tabular}{ c  c  c  c  c  c  c  }
\hline\hline
    & $\overline{x}_i$ $(MW)$ &  $a_i$ $(\$/MW^2)$ & $b_i$ $(\$/MW)$ & $c_i$ $(\$)$\\
\hline
    $DG_1$ & 70 & 0.27 & 40 & 3.4\\
    $DG_2$ & 60 & 0.3 & 26.5 & 3\\
\hline\hline
\end{tabular}}
\end{center}
\end{table}

\begin{table}[h]
\caption{Regulation costs}\label{Table 9}
\begin{center}
\renewcommand\arraystretch{1.5}{
\begin{tabular}{  c  c c c}
\hline\hline
       $c_{D,1}$/\$ & $c_{D,2}$/\$ & $c_{U,1}$/\$ & $c_{U,2}$/\$\\
\hline
    10  &  20 & 100 & 
    200\\
\hline\hline
\end{tabular}}
\end{center}
\end{table}

\begin{table}[h]
\caption{Technical data of real-time operation}\label{Table 10}
\begin{center}
\renewcommand\arraystretch{1.5}{
\begin{tabular}{  c  c c c}
\hline\hline
       $\bar{z}_1^D$/MW & $\bar{z}_2^D$/MW & $\bar{z}_1^U$/MW & $\bar{z}_2^U$/MW\\
\hline
    10  &  30 & 10 & 
    30\\
\hline\hline
\end{tabular}}
\end{center}
\end{table}

\section{The Summary of Quantile Predictor and Agent's Parameters}\label{Appendix B}

\begin{table}[h]
\caption{Summary of QMLP parameters}\label{Table 1}
\begin{center}
\begin{tabular}{ c  c }
\hline\hline
    Item & Value\\
\hline
    Batch size & 128\\
    No. of hidden layers & 2\\
    No. of neurons in input layer & 4\\
    No. of neurons in output layer & 1\\
    No. of neurons in hidden layer & 128\\
    Optimizer & Adam\\
    Learning rate & 1e-3\\
\hline\hline
\end{tabular}
\end{center}
\end{table}

\begin{table}[h]
\caption{Summary of Agent Parameters}\label{Table 2}
\begin{center}
\begin{tabular}{ c  c }
\hline\hline
    Item & Value\\
\hline
    Batch size & 128\\
    No. of hidden layers & 2\\
    No. of neurons in the first hidden layer & 512\\
    No. of neurons in the second hidden layer & 256\\
    No. of neurons in input layer & 4\\
    Optimizer & Adam\\
    Learning rate & 1e-4\\
\hline\hline
\end{tabular}
\end{center}
\end{table}





\end{document}